\documentclass[reprint,aps,pra,twocolumn,showpacs ,superscriptaddress,groupedaddress, floatfix]{revtex4-1} 
\usepackage{hyperref}

\hypersetup{colorlinks=true, citecolor=blue, urlcolor=blue, linkcolor=blue}

\usepackage{graphicx}  
\usepackage{dcolumn}   
\usepackage{bm}        
\usepackage{bbold}
\usepackage{appendix}
\usepackage{amssymb, amsmath}   
\usepackage[usenames, dvipsnames]{color}
\usepackage{float}

\usepackage[latin1]{inputenc}

\begin{document}
\title{Multi-level Purcell effect and the impact of vibrational modes in molecular quantum optics}
\author{Charlie Nation}%
\email{C.Nation@ucl.ac.uk}
\author{Valentina Notararigo}%
\author{Alexandra Olaya-Castro}%
\affiliation{Department of Physics and Astronomy, University College London, London WC1E 6BT, United Kingdom}    

\date{\today}\label{key}

\begin{abstract}
 The increased decay rate of a two-level system weakly coupled to an optical cavity, known as the Purcell effect, is a cornerstone of cavity QED. However, the effect of cavity coupling is not well understood if the two-level system is replaced by a multi-level interacting system. Motivated by experiments looking to characterise molecular systems via exploiting a cavity interaction, we study a manifestation of the Purcell effect in a bio-inspired photosynthetic dimer. We focus in particular on how molecular vibrational modes, thought to play an important role in photosynthetic exciton transport, impact the system-cavity behaviour in the Purcell regime. We provide a theoretical picture in terms of an effective non-Hermitian Hamiltonian, which extends the simple picture of a Jaynes-Cummings model to the description of a `multi-level' Purcell effect, where different levels have differing Purcell factors, with effective cooperativities mediated by coherent vibrational interactions.
\end{abstract}

\maketitle
\section{Introduction}

Molecular systems often exist at the boundary of quantum and classical phenomena. The coherent coupling of electronic, optical, and vibrational degrees of freedom available in molecular systems thereby offers an excellent testbed for optomechanical and quantum optical effects via molecular cavity quantum electrodynamics (QED) experiments \cite{Flick2017, Ojambati2019, Wang2021}. Of particular interest to study quantum effects in photosynthetic molecules is the detailed interplay between coherent and dissipative interactions, and the effect of bath degrees of freedom in the form of vibrational modes on exciton transport \cite{Mirkovic2007a, Olaya-Castro2011, Oreilly2014, Dean2016, Duan2017, Caycedo-Soler2018, calderon2023}.
More generally, many-body system-cavity interactions are of relevance to fields such as ultracold atomic gases \cite{Chiacchio2019}, as well as optical properties of quantum-dots \cite{Laussy2009, DelValle2009, Nabiev2010, Laussy2011, Muller2015, Tufarelli2020}, photonic devices \cite{Raftery2014, Fernandez-Lorenzo2021}, condensed-matter physics \cite{phillips2020}, artificial light-harvesting devices \cite{Romero2017}, and a variety of other applications in quantum technologies, quantum chemistry, and beyond.

Central to emitter-photon interactions via cavity QED is the Purcell effect \cite{Purcell1946}, canonically modelled by the Jaynes-Cummings (JC) model, whereby two level system weakly interacting with a `bad' cavity (cavity decay $\kappa$ is fast compared to its resonant frequency $\omega_c$, characterised by the Quality factor $Q := \frac{\omega_c}{\kappa}$) experiences a significantly increased decay rate compared to its bare emission rate $\gamma$. This increase is determined by the cavity cooperativity $C = g_c^2 / \kappa\gamma$, with $g_c$ the system-cavity coupling strength.
Molecule and many-atom cavity interactions have been theoretically described via modifications to canonical models of cavity QED \cite{Cwik2016, Plankensteiner2019b, Tufarelli2020, Zhang2021}. 
Here we describe the coherent electronic and exciton-vibration interactions in multi-level molecular systems, and their manifestation in the Purcell enhancement of the decay of molecular states. 
A crucial motivation for analysis of the molecular Purcell effect is that it may be exploited in order to increase light capture for quantum optical experiments \cite{Bujak2011, Sapienza2015, Trojak2020}, and is thus of particular relevance for single molecule spectroscopies, where a signal may otherwise be extremely weak.

Experimental application of fluorescence enhancement at the single molecule level has previously been exploited in order to measure photon statistics in the photosynthetic LH2 complex \cite{Wientjes2014}, showing antibunching of emitted light. Both experimental \cite{Coles2014a, Konrad2014} and theoretical \cite{Caruso2012, Saez-Blazquez2019, Zhao2020} studies of photosynthetic molecules interacting with a cavity have mostly concerned strong-coupling regimes, polariton formation, and the optimisation or modification of transport. Notably, in Ref. \cite{Wang2019} targeted Purcell enhancement on a molecular transition was used in order to turn a molecule into an ideal quantum emitter. This experiment captures one branch of the possible applications of molecular cavity QED: creation of a hybrid system with a new behaviour induced by cavity interaction. Indeed, the Purcell effect may be exploited to selectively effect particular transitions of molecular systems in this manner \cite{Cang2013, Metzger2019, Wang2021}. In this work we are motivated by a second branch: exploiting a cavity coupling in order to \emph{probe} molecular behaviour, with no (or at least controlled) effect on the molecular system itself. This enables the cavity to act as a tool for increased targeted light collection, which is otherwise extremely weak in the single molecule regime.

In Ref. \cite{Caruso2012} Caruso et. al. show that the emission spectra from the cavity coupled to LH2 yields information on the delocalisation of excitonic states. Here we study a similar scenario, probing in which limits the resulting enhanced light emission may faithfully reflect the properties of a bare molecule, providing a theoretical description of the effect of the cavity mode on a multi-level molecule, extending the focus to the role of vibrational coherences.

We take as our example model a prototype photosynthetic dimer \cite{Holdaway2018} including coherent vibronic couplings and incoherent environmental interactions. We find that even in the weak cavity coupling regime an understanding of the cavity coupling in terms of the Purcell effect must be modified. Whilst we may indeed understand the system in terms of modified decay rates of excited states, in general these decay rates are effected differently for different, near resonant, molecular states. We give a simplified analytical description of these differing Purcell factors in terms of an effective non-Hermitian Hamiltonian, and see that Jaynes-Cummings like couplings emerge, characterised by state dependent cavity cooperativities, which depend crucially on both excitonic and vibrational coherences of the system.

This article is arranged as follows. We begin in Section \ref{sec:JC} by introducing the Purcell effect and its analysis via an effective non-Hermitian Hamiltonian with the simplest case of the Jaynes-Cummings (JC) model. In Section \ref{sec:dimer_intro} we extend the analysis to a multi-level system. We first introduce the prototype photosynthetic dimer model, in which we show that the Purcell effect can be observed directly from excited state dynamics in \ref{sec:dynamics}. We then in Section \ref{sec:coops} show that this direct approach is somewhat naive, and in terms of an effective Hamiltonian in analogy to the JC model analysis, calculate effective Purcell factors for the system, and observe that they in general differ for different (near resonant) states. In Section \ref{sec:vib} we analyse in more detail how molecular vibronic and excitonic delocalisations play a role in the MLP effect. We finally conclude in Section \ref{sec:conclusion}. Additional derivations are provided in the appendix.

\section{The Purcell effect}\label{sec:JC}

\begin{figure}
	\includegraphics[width=0.45\textwidth]{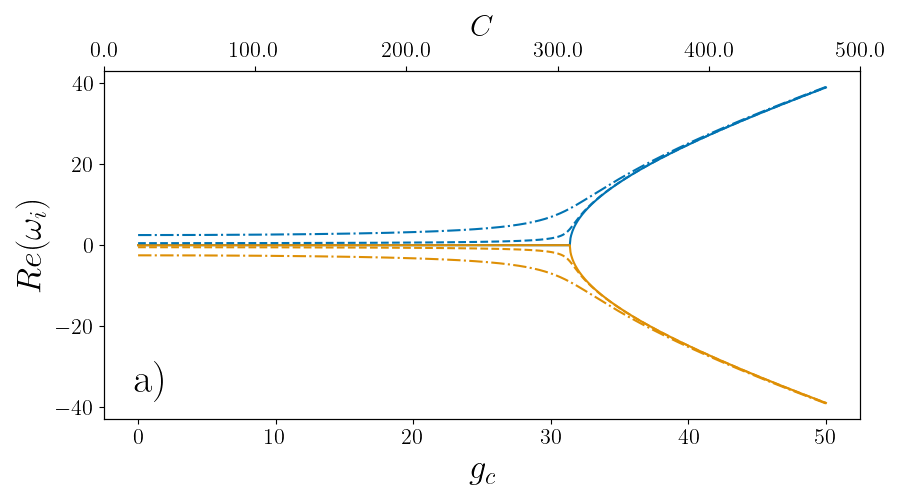}\\
	\includegraphics[width=0.45\textwidth]{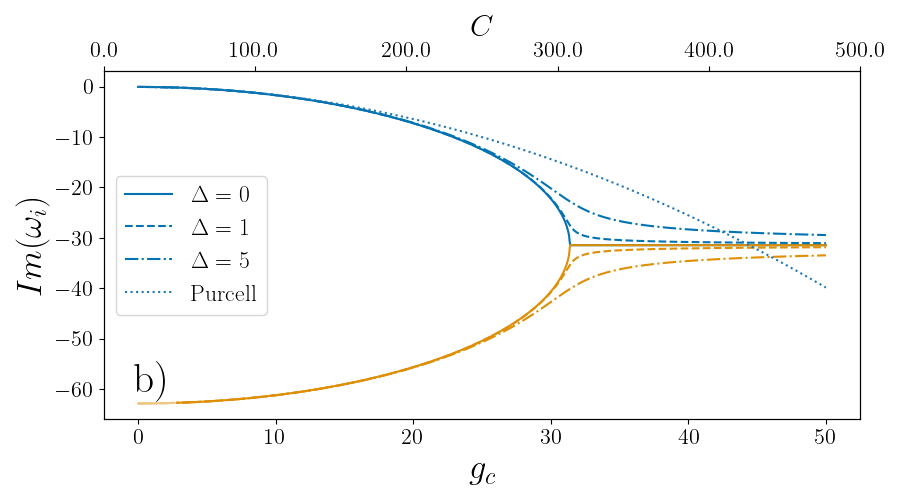} 
	\caption{a) Real and b) imaginary parts of the eigenspectrum of the effective non-Hermitian Hamiltonian of the Jaynes-Cummings model, Eq. \eqref{eq:JCeffective_hamil}, for varying detuning $\Delta := \omega_0 - \omega$. Blue lines show dressed atom modes, and orange lines show dressed cavity modes. Dotted blue line shows Purcell enhancement of the decay rate via $\gamma \to \gamma(1 + 4C)$. Parameters: $\kappa / 2\pi = 20, \, \gamma / 2\pi = 0.02, \omega_0 = 1$.}
	\label{fig:JC_detuning}
\end{figure}

The canonical demonstration of the Purcell effect is via the Jaynes-Cummings (JC) model, describing a two level system interacting with a single cavity mode
\begin{align}\label{eq:H_JC}
H_{JC} = \frac{1}{2}\omega_0 \sigma_z + \omega_c a^\dagger a + g_c \sigma_+ a_k + g_c^*\sigma_- a^\dagger_k.
\end{align}
Cavity and atom decay processes are treated via Markovian dissipators in GKSL form, $\mathcal{D}_L[\cdot] = 2L \cdot L^\dagger - \{L^\dagger L, \cdot\}$, for some jump operator $L$. These are taken to be the atom and cavity annihilation operators $\sigma_-$ and $a$, respectively. The master equation governing the atom-cavity system evolution is thus,
\begin{align}
\partial_t \rho(t) = -i[H_{JC}, \rho(t)] - \frac{\gamma}{2} \mathcal{D}_{\sigma_-}[\rho(t)] - \frac{\kappa}{2} \mathcal{D}_{a}[\rho(t)].
\end{align}
In order to describe the Purcell effect, it is useful to re-express the master equation in terms of an effective non-Hermitian Hamiltonian \cite{Lien2016}. For a GKSL master equation this takes the form \cite{Choi2010}
\begin{align}\label{eq:effective_hamil}
H_{eff} = H - i\sum_i \frac{\gamma_i}{2} L^\dagger_i L_i ,
\end{align}
where $H$ is the original Hamiltonian describing unitary coherent dynamics, and the second term manifests the effect of the Markovian decay channels. We thus obtain,
\begin{align}\label{eq:JCeffective_hamil}
H_{eff} = H_{JC} - i\frac{\gamma}{2} \sigma^\dagger \sigma - i\frac{\kappa}{2} a^\dagger a.
\end{align}
This effective Hamiltonian is easily diagonalised, from which we obtain complex eigenvalues for the lowest energy doublet given by
\begin{align}
\omega_\pm = \frac{\omega_c + \omega_0 +i(\kappa + \gamma)}{2} \pm \sqrt{g_c^2 + \left( \frac{\Delta - i(\kappa - \gamma)}{2}\right)^2},
\end{align}
where $\Delta = \omega_c - \omega_0$ is the cavity detuning.

Importantly for the following analysis, the imaginary parts of the complex eigenvalues may be understood as the decay rate of the corresponding Hamiltonian (pseudo) eigenstate, which can be seen by writing explicitly the evolution of some pure state $|\phi\rangle$ via $|\phi(t)\rangle = e^{-iH_{eff}t}|\phi\rangle = \sum_{\mu} e^{-iE_\mu} c_\mu |\psi_\mu\rangle $. Here $|\psi_\mu\rangle$ are the eigenstates of $H_{eff}$, and $c_\mu = \langle \psi_\mu|\phi\rangle$. We see, then, that for complex $E_\mu$ individual terms oscillate at a frequency $\mathcal{R}e[E_\mu]$, and decay at a rate $\mathcal{I}m[E_\mu]$.

In Fig. \ref{fig:JC_detuning} we show the dependence of the real and imaginary parts of the effective Hamiltonian eigenvalues on the cavity coupling strength $g_c$ for a given quality factor. For $g_c = 0$, the eigenstates correspond to uncoupled atom and cavity modes, decaying at rates $\gamma$ and $\kappa$ respectively. As the coupling is increased these modes are hybridised. The Purcell regime is characterised by the area in which the respective decay rates are altered by the interaction, but the energy levels remain largely unaffected.
This change in decay rate is determined by cavity cooperativity \cite{Carmichael2},
\begin{align}\label{eq:cooperativity}
C := \frac{g_c^2}{\kappa\gamma},
\end{align}
through the modification to the bare atom decay rate via $\gamma \to \gamma(1 + 4C)$. This is shown on Figure \ref{fig:JC_detuning}b) (dotted line), and compared to the decay rates obtained directly through diagonalization of the effective Hamiltonian. We see that the Purcell regime (where the above prescription is a good approximation) extends into rather high cooperativities of $C \lesssim 200$, corresponding to a significant increase in the light emission from the atom. We additionally see in Figure \ref{fig:JC_detuning} that this picture holds upon introducing a detuning \cite{Carmichael2}. The pertinent question for application to molecular cavity QED, then, is whether, and in what limits, this picture can be extended to interacting multi-level systems coupled to a cavity mode.

\section{Multi-Level Purcell Effect}\label{sec:MLP}

\subsection{Prototype Photosynthetic Dimer}\label{sec:dimer_intro}

For our analysis, we use a biologically inspired dimer model \cite{Oreilly2014, calderon2023, chuang2023}, which consists of two pigments, and their surrounding Bosonic environment. The electronic degrees of freedom of the pigments are each described by a two-level system,
\begin{align}
H_{el} = \epsilon_1 \hat{n}_1 + \epsilon_2 \hat{n}_2 + V(\sigma_1^\dagger \sigma_2 + \sigma_2^\dagger \sigma_1)
\end{align}
where $\sigma_k = |G\rangle \langle k|$, and $\hat{n}^{(k)} = \sigma_k^\dagger \sigma_k$ for each pigment site $k$. We thus have that the excited states of each site are coupled by $V$, and differ in energy by $\Delta \epsilon = |\epsilon_1 - \epsilon_2|$. The ground state $|G\rangle$ and doubly excited state $|1, 2\rangle$ are each electronically uncoupled. The central dimer of excited states can then be diagonalised, transforming to the excitonic basis, $|X_1\rangle, |X_2\rangle$ have energies
 $E + \frac{\Delta E}{2}$ and
 $E - \frac{\Delta E}{2}$, with
$\Delta E = \sqrt{\Delta \epsilon^2 + 4V^2}$.
 Note that the transformation to the excitonic basis does not affect the ground or doubly excited states.

The role of electronic interaction $V$ in delocalising the excitons may be characterised by a `mixing angle' $\theta$, via
\begin{align}\label{eq:theta_def}
\zeta = \tan(2\theta) = \frac{2V}{\Delta \epsilon}.
\end{align}
The excitons are then simply written in terms of the localised pigment excitations via $|X_1\rangle = \cos(\theta)|2\rangle - \sin(\theta)|1\rangle $, $|X_2\rangle = \cos(\theta) |1\rangle + \sin (\theta) |2\rangle.$

The role of excitonic delocalisation in quantum transport is of particular interest to the study of photosynthetic molecules, and is captured for this model by $\zeta$. We choose parameters to resemble the cryptophyte antennae PE545 \cite{Curutchet2013}, which are detailed in Table \ref{table:dimer_params}.

\begin{table}
	\begin{tabular}{ |c|c| } 
		\hline 
		$\zeta = \frac{2V}{\Delta\alpha}$ & 0.1  \\ 
		$V$ & 92cm${}^{-1}$  \\ 
		$\Delta\alpha$ & 1042cm${}^{-1}$  \\ 
		$\Delta E$ & 1058.2cm${}^{-1}$  \\ 
		$\gamma$ & (0.5ns)${}^{-1} \approx $ 0.01 cm$^{-1}$  \\
		$E$ & 18000cm${}^{-1}$  \\
		$\gamma_{pd} $ & (1ps)${}^{-1} \approx$ 5.31 cm$^{-1}$ \\
		$\omega_{vib} $ & 1111 cm${}^{-1}$  \\
		$P_{X_1}$ & (0.6ns)${}^{-1} \approx$ 0.009 cm$^{-1}$ \\
		$g$ & 267.1cm${}^{-1}$  \\
		 $\Gamma_{th}$ & (1ps)${}^{-1} \approx$ 5.31 cm$^{-1}$ \\
		$\beta $& $(K_BT)^{-1} = 300$K  \\
		 $\Gamma$ & (0.48ps)${}^{-1} \approx$ 1.11 cm$^{-1}$ \\
		\hline
	\end{tabular}
	\caption{\label{table:dimer_params} Parameters of Dimer model, chosen to resemble cryptophyte antennae PE545 \cite{Curutchet2013}. }
\end{table}

We additionally include coherently coupled vibrational modes, which have energies taken near resonance to the excitonic energy gap. Such near resonant vibrational modes are understood to be a key potential mechanism contributing to the emergence of long-lived coherences in photosynthetic complexes \cite{Olaya-Castro2011, Huelga2013, Oreilly2014, Nalbach2015, Dean2016, Blau2018a, Caycedo-Soler2018, Bennett2018, Higgins2021}. Our phenomenological model may be understood as absorbing the non-Markovian contribution of the environment into the coherent dynamics, and thus allowing the additional environmental effects of the bath to be described by Markovian dynamics. This is often achievable explicitly via the reaction coordinate picture \cite{Nazir2018, Correa2019}.

The Hamiltonian describing the vibrational mode is $H_{vib} = \omega_{vib} (d_1^\dagger d_1 + d_2^\dagger d_2)$, which is coupled to the electronic states via,:
\begin{align}\label{eq:H_el_vib}
H_{el-vib} =  g\sum_{k=1}^2 \sigma^\dagger_k \sigma_k (d^\dagger_k + d_k),
\end{align}
where $d_k$ are the vibronic annihilation operators on sites $k$, and have energies $\omega_k$.

Additionally, we wish to capture the interaction of the above system with a single mode optical cavity with creation (annihilation) operators are given by $b^\dagger, \, (b)$, with Hamiltonian $H_c = \omega_c b^\dagger b$. We make the rotating wave approximation, thus assuming that the coupling to the cavity $g_c$ is weak relative to the electronic coupling $V$, and thus have a Tavis-Cummings like interaction
\begin{align}\label{H_el_c}
H_{el-c} =  \frac{g_c}{2}\sum_{k}^2  [\sigma_k b^\dagger + \sigma_k^\dagger b].
\end{align}
We choose the cavity frequency to be fixed near resonance with the highest energy exciton.

We describe the environment via a sum of the various incoherent processes acting on the system. We apply a pure dephasing in the site basis at a rate $\gamma_1 = \gamma_{pd}$, via jump operators $\mathcal{L}_{1,k} = A_k$, with $A_k = |k\rangle \langle k|$, for $k = 1, 2$. Additionally, we account for thermal relaxation and absorption at rates $\gamma_2 = \Gamma_{th}(n(\omega_{vib}) + 1)$ and $\gamma_3 = \Gamma_{th}n(\omega_{vib})$ respectively, via the jump operators $\mathcal{L}_{2, k} = d_k$ and $\mathcal{L}_{3,k} = d^\dagger_k$. We further model the radiative decay of polaritonic states (the eigenstates of the coupled exciton-vibration-cavity system), denoted $|F_\nu\rangle$, into the vacuum dictated by rate $\gamma_4 = \gamma$, with state-dependent rates $\gamma_{\nu, l} = F_{\nu, l} \gamma$, where $F_{\nu, l} = \sum_m \langle m, l| F_\nu \rangle$ dictates the overlap between excited polaritonic states and vibrational excitations in the electronic ground state, and with associated jump operators $\sigma_{v, l}=|G, l\rangle \langle F_v|$. This form for the jump operators describing decay processes via the dipole operator is derived in Appendix \ref{app:jump_ops}.
Additionally, we include a weak incoherent pumping of the highest energy exciton at rate $\gamma_5 = P_{X_1}$ via the jump operator $\mathcal{L}_5 = \sigma^\dagger_{X_1} = |X_1\rangle\langle G|$, and a decay of the cavity mode at rate $\gamma_6 = \gamma_c$ via jump operator $\mathcal{L}_6 = b$.

In summary, the set of jump operators we consider are: $\{\mathcal{L}_i\} = \{A_k, d_k, d_k^\dagger, \sigma_{vl}, \sigma_{X_1}^\dagger, b\}$, with respective rates $\{\gamma_i\} = \{\gamma_{pd},\, \Gamma_{th}(n(\omega_{vib}) + 1),\, \Gamma_{th}n(\omega_{vib}),\, \gamma_{\nu, l} ,\, P_{X_1},\, \gamma_c\}$. 

\subsection{Dynamics}\label{sec:dynamics}
\begin{figure}
	\includegraphics[width=0.45\textwidth]{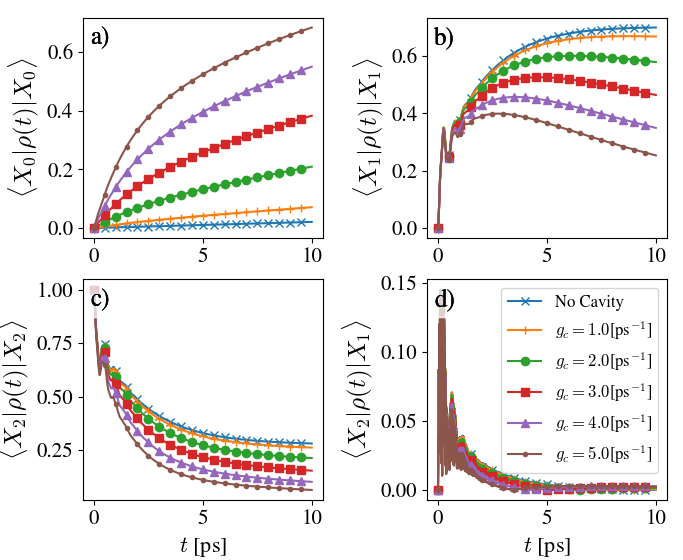}\\
	\includegraphics[width=0.45\textwidth]{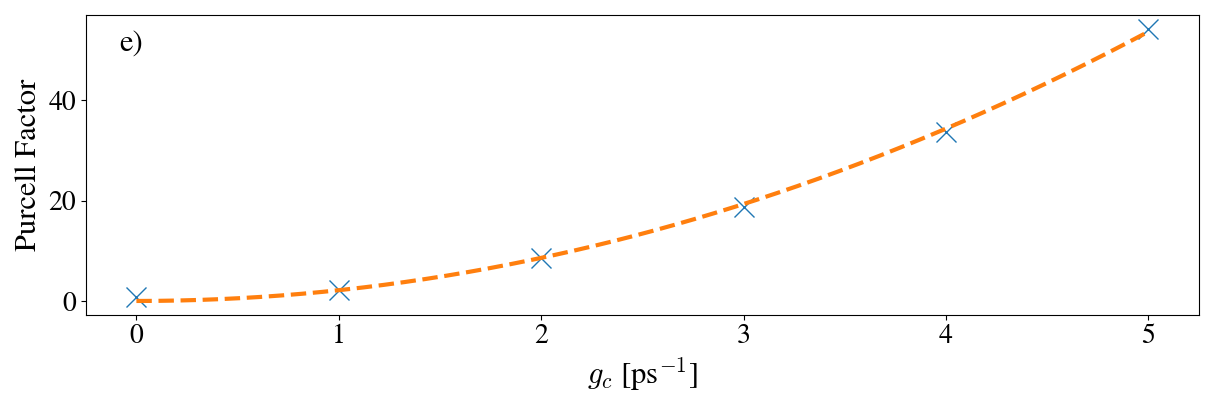}
	\caption{Dynamics of density operator matrix elements $\rho_{|X_i\rangle\langle X_j|}(t)$ corresponding to excitonic populations ($i=j$, a) - c)) and coherences ($i\neq j$, d)) for varying $g_c$. Here we observe the increased rate of decay of excited state populations and coherences due to the Purcell effect. Purcell factor as obtained via a fit to the exponential growth of ground state populations (a)) are shown in e). Fit to $(1 - e^{- \gamma^\prime t})$ where $\gamma^\prime = \gamma_0(1 +F)$, and $F$ is the Purcell factor extracted from fit, and $\gamma_0$ is the bare growth rate with no cavity present. $Q = 50$, $\beta = 300$K.}
	\label{fig:dynamics}
\end{figure}

The Purcell effect is observed most readily as an increase in the decay rate of the excited state of a system. For a many-body system, however, there are additional timescales that play a role in the dynamics. One can see, therefore, that a complete characterisation of the Purcell regime becomes less clear in a molecular-cavity QED setup, where internal system dynamics and system-cavity dynamics may be convoluted.

In order to observe this interplay of system and decay timescales, in Fig. \ref{fig:dynamics} we show the dynamics of relevant density operator matrix elements for varying $g_c$. We initialise the system in the state $\rho(0) = |X_1\rangle \langle X_1|\otimes \rho_{th}^{(vib)} \otimes |0\rangle_c {}_c \langle 0|$, where $\rho_{th}^{(vib)} := \frac{1}{Z}e^{-\beta H_{vib}}$ is the thermal state of the vibrational Hamiltonian, and $|0\rangle_c$ is the cavity ground state. Notice that in the free dimer model (no cavity) we see a clear separation of the timescales of internal system dynamics and the decay to the ground state (the latter occurring on longer timescales than those shown), yet as $g_c$ is increased we can observe the Purcell effect in the increased decay rate of the excited states. As expected, this increase is observed to be of the form $\gamma \to \gamma(1 + F)$, where $F \propto g_c^2$ (Figure (\ref{fig:dynamics}e)).

One may then interpret the `Purcell regime' in two ways: Firstly, simply as the increase of the rate of decay of the molecule, with no regard to the effect on the internal dynamics, or second, in terms of the effect on the rate of decay of individual states within the system, enforcing each level is itself in the Purcell limit. In the following, we refer to these regimes as the Jaynes-Cummings-Purcell regime (JCP), and the multi-level Purcell regime (MLP), respectively.

\begin{figure*}
	\includegraphics[width=0.45\textwidth]{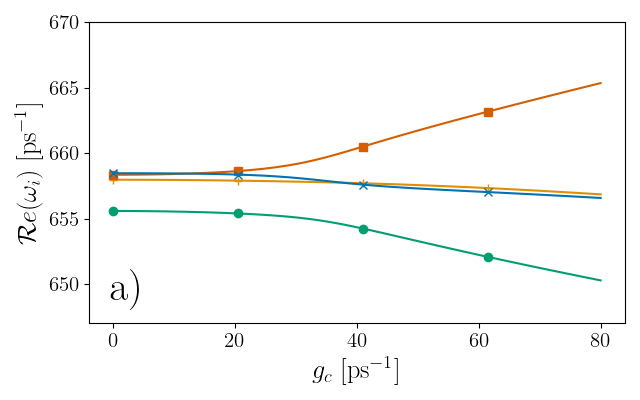}
	\includegraphics[width=0.45\textwidth]{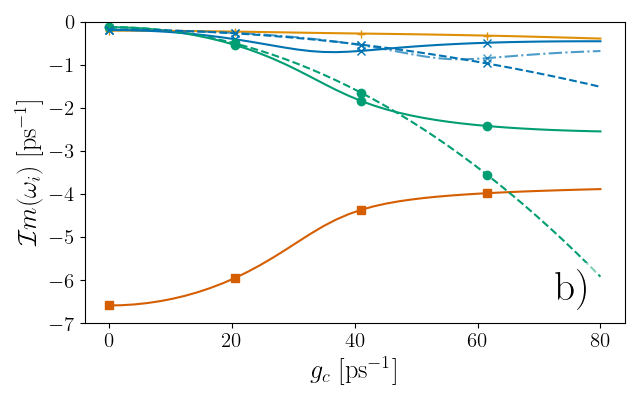}
	\caption{(Colour online) a) Real and b) imaginary parts of the eigenspectrum of the effective non-Hermitian Hamiltonian of the dimer model. Shows only the single excitation manifold of states corresponding to  $|X_1, 0, 0, 0\rangle$ (green circles), $|X_2, 1, 0, 0\rangle$ (blue crosses),  $|G, 0, 0, 1\rangle$ (red squares) and  $|X_2, 0, 1, 0\rangle$ (yellow pluses) at $g_c = 0$. Solid lines show exact calculation of energy levels. Dotted lines in b) show Purcell regime estimates for decay rates from Eq. \eqref{eq:DimerCoop}, assuming no higher vibrational levels. Shown for biological value $g_v = 267.1$ cm$^{-1}$, where we can see deviation due to the role of higher excited states via blue dash-dotted line in b), showing imaginary decay rate of  $|X_2, 1, 0, 0\rangle$ for restricted vibrational excitation number to $L_v = 1$. We can see that the centre of mass mode branch changes very little with increasing cavity coupling, indicating its effective decoupling over the Purcell regimes of other excited molecular states. $Q=50$.}
	\label{fig:dimer_NH_full}
\end{figure*}

Obviously, the latter condition is a much stronger requirement, and we may not expect such a regime in systems where the internal dynamics occurs on a timescale comparable with its excited state decay in free space, as any change to this decay rate necessarily alters the internal dynamics. However, it is commonplace in many chemical and biological systems of interest for these timescales to be well separated \footnote{A clear separation of these timescales is implicit in typical approaches, for example, to calculate fluorescence spectra of molecular systems \cite{Mukamel}, where it is assumed that the decay of the system into the electromagnetic field occurs from an internal thermal state - that is, after the internal dynamics has reached its steady state, but before photon emission.}. For example, in the model described here, typical of many photosynthetic pigment protein complexes, internal dynamics occur over a few picoseconds, whereas decay to the ground state occurs over $\sim 1$ ns. One may thus expect that the Purcell effect may enable a significant increase in the decay rate without a large effect on internal dynamics.

Indeed, in Figure \ref{fig:dynamics}e), then, we see that the JCP regime is observed in the prototype photosynthetic dimer introduced above. In what follows, we apply a similar effective Hamiltonian description to the JC model analysis above in order to characterise the effect of internal electronic and vibrational degrees of freedom on the Purcell effect, enabling characterisation of the MLP regime.

\subsection{Effective cooperativities}\label{sec:coops}

The effective Hamiltonian of the photosynthetic dimer model may be obtained from Eq. \eqref{eq:effective_hamil}, as in the case of the JC model.
In Figure \ref{fig:dimer_NH_full}, we show the change in the real and imaginary parts of selected eigenenergies (see below) with cavity coupling $g_c$. What is immediately apparent is phenomenological similarity to the JC model. For weak couplings we have a very small change to the energy levels of the system (Figure. \ref{fig:dimer_NH_full}a)), complemented by comparatively large changes to the decay rates (Figure. \ref{fig:dimer_NH_full}b)).

As with the JC model, the energy levels $\mathcal{R}e(\omega_i)$ are well separated into bands corresponding to manifolds of like exciton + cavity occupation number, which for weak couplings are not significantly mixed by the presence of the cavity.
Of particular relevance to spectroscopic experiments is the single excitation manifold, which is that accessed in the limit of weak pumping. Thus, we attempt to simplify the problem by restricting our regime of interest to the relevant subspace defining this initial band of low energy excitations. The four states which encompass the relevant occupied states in this limit are those with real and imaginary eigenenergies depicted in Figure \ref{fig:dimer_NH_full}.

In Appendix \ref{app:reduced_subspace}, we derive the restricted effective Hamiltonian for the low energy manifold. We re-express the vibrational modes into their relative displacement and centre of mass modes, where the latter is seen to cause no electronic transitions, rather only an energy shift, which we treat in the mean-field. This state is seen in yellow (+ symbols) in Figure \ref{fig:dimer_NH_full}, where the eigenergies are observed to depend extremely weakly on $g_c$, indicating an approximate decoupling from the cavity. We then obtain an effective description of the weak pumping regime in terms of three coupled states, described by the cavity mode $|G, 0_{rd}, 1_{cav}\rangle$, the highest exciton, $|X_1, 0_{rd}, 0_{cav}\rangle$ (labelled (1) below), and the lower exciton with a single vibrational quanta in the relative displacement mode  $|X_2, 1_{rd}, 0_{cav}\rangle$ (labelled (2)). Due to the near resonance of the vibrational mode, with the excitonic energy gap, or small $\Delta_{vib} := \Delta E - \omega_{vib}$, and the tuning of the cavity to near resonance with the highest energy exciton, each of these states is nearby in energy. 

\begin{figure*}
	\includegraphics[width=0.95\textwidth]{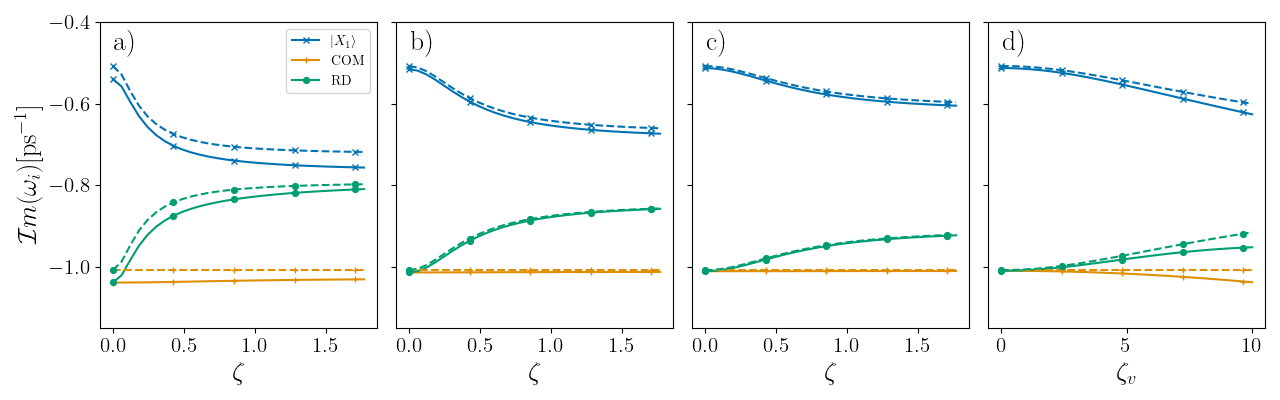}
	\caption{a-c) Imaginary part of eigenenergies $\omega_i$ of Dimer model for varying excitonic delocalisation (see Eq. \eqref{eq:theta_def}) for a) $g= 267.1$ cm$^{-1}$ ($g>V$), b) $g = 100$ cm$^{-1}$ ($g\sim V$) and c) $g = 50$ cm$^{-1}$ ($g<V)$. We see that the reduced model (dashed lines) is in good agreement for vibrational couplings that are $\lesssim V \sim 92$ cm$^{-1}$, and numerical results (solid lines) for full model begin to deviate from the analytical calculation around the biological value, however remains a good phenomenological agreement. d) Dependence of imaginary part of eigenenergies $\omega_i$ of Eq. \eqref{eq:DimerCoop} on vibronic delocalisation for $\zeta = 0.1$. Labels $|X_1\rangle$, RD (relative displacement), and COM (center of mass) refer to the (quasi-)eigenstates of $H$ corresponding to $|X_1, 0, 0, 0\rangle, |X_2, 1, 0, 0\rangle$ and $|X_2, 0, 1, 0\rangle$, respectively, for $g_c$ = 0. We can see that the COM dominated eigenstate is essentially unchanged in the parameter regimes studied, indicating that it effectively decouples. Here $g_c \approx 2.65$ cm$^{-1}$, $Q = 50$.}
	\label{fig:rates_vs_delocs}
\end{figure*}

As with the JC model, for small $g_c$ the decay rates are separated into two bands - a quickly decaying (large negative values) band of modes with finite cavity occupation, and a slowly decaying (small negative values) band of molecular excited states. As $g_c$ is increased, molecule-cavity hybridisation causes these molecular excited states to have an increased decay rate, as expected. The core deviation from the JC model is that for each of the excited states shown, we observe a different rate of change of decay rate with cavity coupling - indicating state dependent Purcell factors of these near resonant exciton-vibrational states. In simplified 4 level model, these are described by $|v_i\rangle = c_i^{(X_1, 0)}|X_1, 0_{rd}, 0_{cav}\rangle + c_i^{(X_2, 1)}|X_2, 1_{rd}, 0_{cav}\rangle$ for $i \in [1, 2]$, and have associated cooperativites,
\begin{align}\label{eq:DimerCoop}
C_{i} = \frac{g_{c,i}(\theta, \phi)^2}{\gamma_i^\prime \Gamma_c^\prime },
\end{align}
where
\begin{align}\label{eq:g_c1}
g_{c,1}(\theta, \phi) = \frac{g_c}{2}(\cos(\theta) - \sin(\theta)) \cos(\phi)
\end{align}
and
\begin{align}\label{eq:g_c2}
g_{c,2}(\theta, \phi) = \frac{g_c}{2}(\cos(\theta) - \sin(\theta)) \sin(\phi)
\end{align}
with $\theta$ defined in Eq. \ref{eq:theta_def}, and $\phi = \frac{1}{2}\arctan(\frac{2g_x}{\Delta_{vib}})$ defining an analogous vibrational mixing angle, with $g_x = - \frac{g \sin (2\theta)}{\sqrt{2}}$, and $g_{c, 1(2)}(\theta, \phi)$ are the effective coupling strengths of the two states. The expressions for $\gamma^\prime$ and $\Gamma_c^\prime$ are given in Appendix \ref{app:reduced_subspace}, and describe the decay rates of the vibronically dressed excitonic states, and vibronically dressed cavity mode, respectively. The Purcell modification to the excitonic decay rates expected via the prescription $\gamma \to \gamma (1 + 4C)$ is shown on Fig. \ref{fig:dimer_NH_full}b) as dotted lines.

We note that the approximate cooperativities are underestimated compared to the numerical calculation. This deviation is due to the effect ignored additional vibrational levels in the reduced subspace model, which we show in Figure \ref{fig:dimer_NH_full} (dot-dashed blue line), showing the exact numerical calculation with a restricted vibrational cutoff to a single excitation. This is also seen below in Figure \ref{fig:rates_vs_delocs}, where the analytical results are seen to be significantly closer to the exact numerical results for smaller $g_v$, where higher vibrational states are less important. Crucially, however, the approximate analytical result confirms the key mechanism involved in the multi-level Purcell effect as the mixing of nearby energy levels in the system by vibronic interactions.
We observe in numerical calculations a convergence in vibrational occupation cutoff of $n_v = 4$, and $n_c = 2$ for the cavity mode maximal occupation. We see, however, that the core phenomenology is well captured in the approximate model, and will see in the next section that the model similarly captures more subtle effects of electronic and vibrational coherences on cavity cooperativities.

We thus observe that a simplified picture of the system in terms of a restricted set of energy levels captures the key mechanism of the Purcell effect in a multi-level molecular system, which can be understood as the mixing of molecular eigenstates with the quickly decaying cavity modes. This mixing occurs at different rates for different molecular states, and thus multiple Purcell factors are necessary, each dictated by details of the internal molecular interactions and cavity coupling. 

It is important to note that in the above model a treatment of all three levels is vital to a correct characterisation of the cavity cooperativities, as those of each state depend on one another via the influence of the vibrational mode. This mode has the effect of mixing the excitonic states, leading to a competition for cavity cooperativity between the excitonic states, as we study in more detail in the next section.
%

\subsection{Vibronic modification of Purcell factor}\label{sec:vib}

As we have seen, the effective cooperativities of each molecular excited state depend on both the excitonic and vibronic mixing angles, $\theta$ and $\phi$, which in turn depend on their respective delocalisation parameters $\zeta = \frac{2V}{\Delta\epsilon}$ and $\zeta_v = \frac{2 g_x}{\Delta_v}$, with $\Delta_{vib} = \Delta E - \omega_v$.
In this section we study the role of these excitonic and vibronic delocalisations on the cavity cooperativities, in order to understand the role of  internal molecular coherences on the cavity interaction. To do so, we ensure that vibrational resonance condition is maintained; fixing $\Delta E = \sqrt{\Delta \epsilon^2 + 4V^2}$ and $\omega_v$, and varying the excitonic mixing $\zeta$ by consistently altering $V$ and $\Delta \epsilon$.

In Figure \ref{fig:rates_vs_delocs} we show the dependence of the decay rates of molecular excited states with electronic and vibrational delocalisation parameters $\zeta$ and $\zeta_v$, respectively. We can see immediately that the core effect of the delocalisation is to cause the rates to coalesce, as one may expect moving towards the limit of highly delocalised collective excitations. 

We can understand this behaviour in more detail noting that the cooperativities of each state are dominated by the behaviour of the effective cavity couplings in Eqs. \eqref{eq:g_c1} and \eqref{eq:g_c2}, from which we can see that there are two key parameters dictating the dependence of the effective coupling strengths on excitonic delocalisation. i) the dependence of the effective cavity coupling on $\theta$ (and hence on $\zeta$) when no vibrational mode is present, $g_c(\theta) = \frac{g_c}{2}(\sin(\theta) - \cos(\theta))$, ii)  the vibronic mixing angle itself has a dependence on the excitonic delocalisation, $\phi(\theta) = \frac{1}{2}\tan^{-1}(-\frac{\sqrt{2}g \sin(\theta)}{\Delta_{vib}}) \in [-\frac{\pi}{4}, \frac{\pi}{4}]$ (see Eq. \eqref{eq:H_el_vib2}). For the range of electronic delocalisations in \ref{fig:rates_vs_delocs}a-c), then, we have that due to i) the purely excitonic part is a monotonically increasing function of delocalisation, leading to an increasing decay rate. ii) induces a competing effect caused by the vibronic mode is due to the $\phi(\theta)$ dependent terms for each $g_{c, i}$. This manifests in the increased rate of change of decay rates with $\zeta$ observed as the vibronic coupling strength increases in Figure \ref{fig:rates_vs_delocs}a-c). 

\section{Conclusions}\label{sec:conclusion}

The Purcell effect is a foundation of many applications of cavity QED, and promises to hold a similar place in molecular cavity QED, as more complex systems are studied in such setups. One notable motivation, particularly relevant to single molecule experiments, is the potential enhancement of light collection, which may otherwise be extremely weak. Improved light collection may thus allow for the quantum optics of single molecules to be effectively studied experimentally. As the cavity inevitably hybridises with the system under study, potentially modifying internal coherences and dynamics, if one aims to probe molecular properties a detailed understanding of such hybridisation is necessary in order to determine faithful molecular properties. Here we have shown that the most sensitive effect of the cavity is to change the relative decay rates of the molecular states, where multiple Purcell factors alter decay rates of different states, and are mediated by internal molecular interactions.

We have studied a prototype photosynthetic dimer model, which consists of two excited electronic states, each coherently coupled to a localised vibronic mode near resonance with the transition energy between excitonic state energies. This model captures many of the important features of photosynthetic complexes, and allows for study of vibronic and excitonic coherences. In particular, we observe that the relative change in decay rates under Purcell enhancement is mediated by such coherences, enabling information on the quantum behaviour of the biologically inspired dimer to be accessible via cavity couplings.
 
In order to model the effect of interaction with a single mode cavity on the internal molecular energy structure, we study the effective non-Hermitian Hamiltonian. Indeed, for weak couplings, the real part (energies) of the effective eigenvalues remain approximately constant, whilst there is a large change in the imaginary part (decay rates) - indicative of the Purcell regime. For stronger couplings, however, this picture is quickly broken, as the cavity mixes energy levels within the system, causing an alteration to internal molecular dynamics. We introduce a simplified effective Hamiltonian that captures some of the key phenomena of the photosynthetic dimer studied. From this, we are able to isolate the influence of different effective cavity cooperativities on molecular states, this well captures the behaviour of the differing Purcell factors between near resonant molecular states.

We have further seen that as the different Purcell factors are sensitive to internal molecular interactions, the change in relative decay rates of excited states with cavity couplings may yield information on molecular coherence. We observe that excitonic and vibronic delocalisations each act to alter the molecular cooperativities of particular states of the system. For weak vibronic couplings excitonic delocalisation acts predominantly to increase these cooperativities. However, for larger couplings, as the vibrational and excitonic delocalisations are intimately connected, an increase in the latter causes vibronic couplings between excited states of the system. This has the surprising effect that the cavity cooperativity is not necessarily increasing with excitonic coherence, as, in some regimes, an increase in excitonic delocalisation similarly increases the vibronic mixing on states, decreasing the cooperativity of one state, and increase it for another. This regime coincides with the biologically inspired values of our model, taken to resemble PE545 \cite{Curutchet2013}.

\section{Acknowledgments}

We thank the Engineering and Physical Sciences Research Council (EPSRC UK), and the Gordon and Betty Moore Foundation grant GBMF8820 for financial support. Computer simulations performed with QuTiP \cite{Johansson2012}.

\appendix

\section{Derivation of emission jump operators}\label{app:jump_ops}

In this section we derive the form of the jump operators contributing to the emission processes in the GKSL master equation used in the main text. We use a standard quantum optical master equation \cite{BP_c3}, starting from the light matter interaction Hamiltonian for an $N$-site Frenkel exciton model \cite{Mukamel},
\begin{align}
    H_I = \sum_m \hat{\mu}_m \otimes \hat{B}_m,
\end{align}
with $\hat{B}_m = g_m( \hat{b}^\dagger_m + \hat{b}_m)$ coupled via the dipole operator of site $m$, which may be expressed in the Hamiltonian eigenbasis as follows;
\begin{align}\label{eq:A_omega_def}
    \hat{\mu}_m &= \mu_m (|m\rangle \langle g| + |g\rangle \langle m|) \nonumber \\&
    = \mu_m \sum_{\nu, \nu^\prime} |F_\nu\rangle \langle F_\nu | (|m\rangle \langle g| + |g\rangle \langle m|)|F_{\nu^\prime}\rangle \langle F_{\nu^\prime} | \\&
    = \sum_{\omega} \mu_m(\omega) \hat{\Pi}(\omega)  \nonumber  \\&
    = \sum_{\omega} \hat{A}_m(\omega) \nonumber
\end{align}
where we have defined $\mu_m(\omega) := \mu_m \langle F_\nu | (|m\rangle \langle g| + |g\rangle \langle m|)|F_{\nu^\prime}\rangle$ and $\hat{\Pi}(\omega) =|F_\nu\rangle \langle F_{\nu^\prime} |$, with $\omega = E_\nu - E_{\nu^\prime}$ as the energy difference of the transition $|F_\nu \rangle \to |F_{\nu^\prime}\rangle$. We note that whilst $\hat{\mu}_m$ is Hermitian, the components $\hat{A}_m(\omega)$ are not in general, and that as $|F_\nu\rangle$ are eigenstates of $H$ with energy $E_\nu$, we have
\begin{align}
    [H, \hat{A}(\omega)] = - \omega \hat{A}(\omega), \,  [H, \hat{A}(-\omega)] = \omega \hat{A}(\omega).
\end{align}

Then, after the Born-Markov and rotating wave approximations \cite{BP_c3}, we can write the GKSL form dissipator due to this interaction in the form,
\begin{align}\label{eq:qo_me}
    \mathcal{D}[\rho] = \sum_{\omega}\sum_{m, m^\prime} \gamma_{m, m^\prime}(\omega) (&\hat{A}_m(\omega) \rho \hat{A}^\dagger_{m^\prime}(\omega) \\& \nonumber + \{\hat{A}^\dagger_{m^\prime}(\omega) \hat{A}_m(\omega) , \rho\}),
\end{align}
with 
\begin{align}\label{eq:gamma_def}
    \gamma_{m, m^\prime}(\omega) = \int_{-\infty}^\infty ds e^{i\omega s} \langle \hat{B}^\dagger_m(s) \hat{B}_m(0) \rangle,
\end{align}
Following Ref. \cite{BP_c3} (see sec 3.4), and assuming the electromagnetic environment to be in the limit of small photon number $N(\omega) \ll 1$, we have
\begin{align}\label{eq:qo_me2}
    \mathcal{D}[\rho] = \sum_{\omega>0} \gamma_{m} (\hat{A}_m(\omega) \rho \hat{A}^\dagger_{m}(\omega) + \{\hat{A}^\dagger_{m}(\omega) \hat{A}_m(\omega) , \rho\}),
\end{align}

As we will see, the nature of the dipole operator, enabling transitions from the ground to excited states in the site basis, further restricts the contributing transitions $\Pi(\omega)$.

Before continuing, we note some important features of the relations between the three relevant bases in which processes are here described. First, the site basis $(\{ m \} \, | \, m\in [0, N]) \in \mathcal{B}(\mathcal{H}_{el})$, with $|m=0\rangle = |g\rangle$ being the ground state. The excitonic basis diagonalises the electronic Hamiltonian, which does not couple ground ($m=0$) and excited ($m>0$) states, thus we have similar form of the excitonic basis $(\{ |X_i\rangle \} \, | \, i \in [0, N]) \in \mathcal{B}(\mathcal{H}_{el})$, with $|X_0\rangle = |m=0\rangle = |g\rangle$ labelling the ground state. Finally, we have the total Hamiltonian (polaritonic) eigenstates $|F_\nu\rangle$ which diagonalise the system Hamiltonian including the coherent vibrational modes and cavity modes. We truncate each local vibrational mode at an occupation number $L_v$, and the cavity modes at $L_c$ and thus the total number of vibrational and cavity energy levels is $L_v^2 L_c$, and $\nu \in [0, (N + 1) L^2 L_c]$. In the following we use the multi-index $l$ to refer to both cavity and vibrational mode occupation, that is, $l = (l_{v_1}, l_{v_2}, l_c)$.

Thus, we have
\begin{subequations}
\begin{align}
    &|F_\nu\rangle = \sum_{i, l} c_{i, l}(\nu) |X_i, l\rangle, \\&
    |X_i \rangle = \sum_m a_m(i) |m \rangle,
\end{align}
\end{subequations}

with $c_{il}(\nu) = \langle X_i, l| F_\nu \rangle$ and $a_m(i) = \langle m | X_i \rangle$, with ground state components $c_{0l}(\nu) = \langle g, l| F_\nu \rangle = \delta_{\nu, (0, l)}$ and $a_0(i) = \langle g | X_i \rangle = \delta_{i0}$ respectively. 
Thus, to evaluate the components in Eq. \eqref{eq:A_omega_def}, we can write
\begin{align}
    \langle F_\nu | m \rangle & = \sum_{il} c_{il}(\nu) \langle X_i, l | m\rangle \nonumber \\& 
    =  \sum_{il} c^*_{il}(\nu) a^*_{m}(i) \langle l| 
\end{align}
and
\begin{align}
    \langle F_\nu | g \rangle &= \sum_{il} c_{il}^*(\nu) \langle X_i, l | g\rangle \nonumber \\& 
    = \sum_l  \delta_{\nu, (0, l)} \langle l |
\end{align}
such that
\begin{align}
    \hat{A}_m(\omega) &= \mu_m \langle F_\nu | \left(|m\rangle \langle g| + |g\rangle \langle m|\right)|F_{\nu^\prime}\rangle \hat{\Pi}(\omega)  \nonumber \\&
    =  \mu_m (\langle F_\nu |m\rangle \langle g |F_{\nu^\prime} \rangle + \langle F_\nu|g\rangle \langle m|F_{\nu^\prime} \rangle )\hat{\Pi}(\omega)  \\&
        =  \mu_m (  \sum_{il} c^*_{il}(\nu) a^*_{m}(i) \langle l|  \sum_{l^\prime}  \delta_{\nu^\prime, (0, l^\prime)} |l^\prime \rangle \nonumber \\&
        \qquad \qquad +  \sum_{l}  \delta_{\nu, (0, l)} \langle l | \sum_{il^\prime} c_{il^\prime}(\nu^\prime) a_{m}(i) | l^\prime \rangle  ) \hat{\Pi}(\omega) \nonumber \\&
     =  \mu_m \sum_{il} (  c^*_{il}(\nu) a^*_{m}(i) \delta_{\nu^\prime, (0, l)} + \delta_{\nu, (0, l)}  c_{il}(\nu^\prime) a_{m}(i)  )\hat{\Pi}(\omega) \nonumber 
\end{align}
We thus immediately see that the possible $\omega$ values are those that are close to the ground-excited state transition energy. Concretely, as $\hat{\Pi}(\omega) = (|F_{\nu}\rangle \langle F_{\nu^\prime}|)_{E_\nu - E_\nu^\prime = \omega}$ we have, defining $\hat{\sigma}_{\nu l} := |g, l\rangle \langle F_\nu|$ and  $F_{m l \nu} := \mu_m \sum_i c_{il}(\nu) a_m(i)$,
\begin{align}
    \hat{A}_m(\omega) = \sum_{l} F_{m l \nu} \hat{\sigma}_{\nu l} 
\end{align}
and
\begin{align}
    \hat{A}_m(-\omega) = \hat{A}^\dagger_m(\omega) = \sum_{l} F^*_{m l \nu} \hat{\sigma}_{\nu l}^\dagger 
\end{align}
Thus, we obtain an expression in terms of weighted transitions from Hamiltonian eigenstates to electronic ground states with local vibrational excitations. We assume constant dipole moments $\mu_m = \mu$, and thus write a single rate $\gamma$ for each transition that absorbs the dipole moment contribution to the weights, and define
\begin{align}\label{eq:Aomega}
    A(\omega) &= \sum_m A_m(\omega)  \nonumber \\ &
    = \mu \sum_{m} \sum_l c_{il}(\nu) a_m(i) \hat{\sigma}_{vl}.
\end{align}

In practise it is not convenient to label our set of jump operators by a frequency $\omega$, but rather in terms of quantum numbers that label the unique jump operators. We see in Eq. \eqref{eq:Aomega} that the operators are defined by the indices $\nu, l$, and thus we relabel
\begin{align}
    \sum_\omega A(\omega) = \sum_{\nu} \sum_l F_{\nu, l} \sigma_{\nu, l},
\end{align}
where 
\begin{align}
    F_{\nu, l} &= \sum_m \sum_i c_{il}(\nu) a_m(i) \nonumber \\&
    = \sum_m \sum_i \langle X_i , l | F_\nu\rangle \langle m| X_i\rangle \\&
    = \sum_m \langle m, l | F_\nu\rangle.\nonumber
\end{align}

We can extend the above discussion to include the second excited state noting the additional contribution from $\hat{\mu}_{m, m^\prime} = \mu_{m^\prime} |m, 0\rangle \langle m, m^\prime| + \mu_{m} |m^\prime, m \rangle \langle m, 0| $ for $m \neq m^\prime$. For the dimer model discussed, then, we have two additional processes due to the transitions $|1, 2\rangle \to |m\rangle $ for $m \in [1, 2]$, which have the associated jump operators 
\begin{align}
    \hat{A}_{12 \to m}(\omega) &= \sum_\omega \langle F_\nu |m\rangle \langle 1, 2 |F_{\nu^\prime}\rangle \hat{\Pi}(\omega) \nonumber \\&
    = \sum_\omega \sum_{i, l} c^*_{i, l}(\nu) a^*_m(i) \delta_{\nu, (1, 2, l)} \hat{\Pi}(\omega) \\&
    = \sum_\omega \sum_{i, l} c^*_{i, l}(\nu) a^*_m(i) |X_i\rangle \langle 1, 2|,\nonumber
\end{align}
where we have used that similarly to the ground state contribution the vibrational modes do not mix the single and doubly excited manifolds, and thus $\langle F_\nu| 1, 2\rangle = \delta_{\nu, (1, 2, l)}\langle l|$.

\section{Reduced effective Hamiltonian for prototype photosynthetic dimer model}\label{app:reduced_subspace}
In this section we derive the reduced subspace effective Hamiltonian from which we obtain state-dependent cooperativities. As we are artificially restricting the subspace to that of a single excitation, the form of the Hamiltonian has a non-trivial dependence on the choice of basis in which we truncate, as non-zero couplings to states outside of this subspace remain. In order for the reduced effective Hamiltonian to remain a good approximation, at least for small cavity coupling strengths, we choose a basis that is amenable to such an assumption. 

To this end, we define the relative displacement and centre of mass modes, $d_{rd} = \frac{1}{\sqrt{2}}(d_1 - d_2)$ and $d_{com} = \frac{1}{\sqrt{2}}(d_1 + d_2)$, respectively, and write $|1\rangle = (\cos(\theta) |X_2\rangle - \sin(\theta) |X_1\rangle)$ and $|2\rangle = (\cos(\theta) |X_1\rangle + \sin(\theta) |X_2\rangle)$. The vibrational Hamiltonian then becomes $H_{vib} = \omega_{vib}(d^\dagger_{rd}d_{rd} + d^\dagger_{com}d_{com})$, with a coupling to the electronic degrees of freedom via
\begin{align}\label{eq:H_el_vib2}
H_{el-vib} &= \frac{g}{\sqrt{2}}\sum_{i\in\{1, 2\}} |X_i\rangle \langle X_i| (d_{com}^\dagger + d_{com}) \nonumber \\& 
+ \frac{g}{\sqrt{2}}(\cos (2\theta )\tilde{\sigma}_z- \sin (2\theta )\tilde{\sigma}_x) (d^\dagger_{rd} + d_{rd}),
\end{align}
where the $\tilde{\sigma}_i \, | \, i=\{x, y, z\}$ are the Pauli operators in the excitonic basis. Notice that the centre of mass mode causes only an energy shift proportional to $\frac{g}{\sqrt{2}}$ with respect to the ground state, and therefore has no contribution to coherent excitonic dynamics. Due to the presence of incoherent processes between the excitonic and ground states, however, this energy shift indeed plays a role in when considering decay processes to the ground state.

In this basis, then, we restrict the subspace of interest to those levels around the energy of the highest excitonic state. This using the notation $|X_i\rangle\otimes|n_{rd}\rangle\otimes|n_{com}\rangle\otimes|n_{cav}\rangle = |X_i, n_{rd}, n_{com}, n_{cav}\rangle$, this subspace is then the set of states $\mathcal{H}_1 = \{ |X_2, 0, 0, 0\rangle, |X_1, 1, 0, 0\rangle, |X_1, 0, 1, 0\rangle, |G, 0, 0, 1\rangle \}$. Here we see that, as the vibrational energy $\omega_{vib}$ is near resonant with the difference in excitonic energies, $\omega_{vib} \approx \Delta E$, and further, the cavity is near resonant with the highest level exciton $\omega_{c} \approx E + \frac{\Delta E}{2}$, each of these levels is nearby in energy. 

The cavity coupling Hamiltonian is written in the excitonic basis as
\begin{align}
    H_{el-c} = & \frac{g_c}{2} [(\cos (\theta) - \sin (\theta)|X_1\rangle \langle g| \\& + (\cos (\theta) + \sin (\theta)|X_2\rangle \langle g| ] b + h.c. \nonumber
\end{align}
Note that of the four states in $\mathcal{H}_1$ above, this couples only $|X_1, 0, 0, 0\rangle$ and $|g, 0, 0, 1\rangle$, as additional vibrational transitions are required for the lower excitonic state to interact within the subspace defined.

To obtain the non-Hermitian part we require a similar process. For completeness here we write each of the relevant operators $L^\dagger_i L_i$ in the exciton-rd-com basis as above. $\sum_k A_k^\dagger A_k = \sum_i |X_i\rangle \langle X_i|$, $\sum_k d^\dagger_k d_k = d_{rd}^\dagger d_{rd} + d^\dagger_{com}d_{com}$, $\sum_{k}d_k d^\dagger_k = d_{rd}^\dagger d_{rd} + d^\dagger_{com}d_{com} + 2$. We note that $\sigma_{vl}^\dagger \sigma_{vl} = |F_v\rangle \langle F_v|$, which is already written in the basis diagonalising $H$, and that the jump operators for the cavity mode are unchanged in this basis.


Using the above, then, the real part of the reduced effective Hamiltonian on this single excitation manifold can be obtained from Eq. \eqref{eq:H_el_vib2} as
\begin{widetext}
	\begin{align}\label{eq:H_subspace}
	H_r^{(1)} = \begin{pmatrix}
	E + \frac{\Delta E}{2} & g_x(\theta)  & 0 & g_c(\theta) \\
	g_x(\theta)  & E - \frac{\Delta E}{2} + \omega_{vib} & 0 & 0\\
	0 & 0 &  E - \frac{\Delta E}{2} + \omega_{vib}&  0\\
	g_c(\theta)  & 0 & 0 & \omega_c
	\end{pmatrix},
	\end{align}
	where we have defined $g_x(\theta) = -\frac{g \sin (2\theta)}{\sqrt{2}}$, $g_c(\theta) = \frac{g_c}{2} (\cos(\theta) - \sin(\theta))$, and use the superscript (1) to indicate the subspace of a single excitation. Notice that on this subspace the cavity only couples to the state, $|X_1, 0, 0, 0\rangle$.
	
	From Eq. \eqref{eq:H_subspace} we may immediately notice that the states involving the centre of mass vibrational mode are uncoupled within this subspace. We have observed numerically, however, that ignoring this mode indeed leads to alterations to the observed dynamics and correlation functions. This motivates a mean-field treatment of the centre of mass mode in the reduced subspace, leading to a reorganisation of the exciton mean energy $E \to E_{mf} = E + \frac{g}{\sqrt{2}}\langle X_{com}\rangle$, where $X_{com}$ is the displacement operator for the centre of mass mode, and $\langle \cdots\rangle$ denotes an average in the thermal state of the centre of mass mode. Additionally, the excitonic states are coupled via the relative displacement mode, which is the source of deviation from a pure JC Hamiltonian for this manifold. Notice that the role of the vibrational mode thus depends explicitly on the excitonic delocalisation, which is characterised by the excitonic mixing angle $\theta$.
	
	For $g_c = 0$, in a similar manner, we may diagonalise $H_r^{(1)}$ to obtain the vibronic Hamiltonian, characterised by the vibronic mixing angle $\phi$, where $\tan 2\phi = \frac{2g_{x}}{\Delta_{vib}}$ and $\Delta_{vib} = \Delta E - \omega_{vib}$. Defining $\Delta^\prime = \sqrt{\Delta_{vib}^2 + 4g_x^2}$, the corresponding vibronic eigenstates are

		\begin{align}\label{eq:vibronic_states}
		 &|v_1 \rangle = \left|\sqrt{1 + \frac{\Delta_{vib}}{\Delta^\prime}}, \sqrt{1 - \frac{\Delta_{vib}}{\Delta^\prime}}, 0  \right\rangle = |\cos( \phi), \sin (\phi), 0\rangle \nonumber \\&
		 |v_2 \rangle = \left|-\sqrt{1 - \frac{\Delta_{vib}}{\Delta^\prime}}, \sqrt{1 + \frac{\Delta_{vib}}{\Delta^\prime}}, 0  \right\rangle = |-\sin (\phi), \cos (\phi), 0 \rangle \\&
		 |v_c \rangle = \left|0, 0, 1 \right\rangle \nonumber,
		\end{align}
	with corresponding energies $\epsilon_1 = E_{mf} + \frac{\omega_{vib}}{2} + \Delta^\prime, \epsilon_2 = E_{mf} + \frac{\omega_{vib}}{2} - \Delta^\prime, \epsilon_3 = \omega_c$. Expressed in this vibronic basis, the Hamiltonian then becomes
	
	\begin{align}\label{eq:Hr_sub_vibronic}
	\tilde{H}^{(1)}_r = 
	\begin{pmatrix}
	\epsilon_1 & 0 & g_c(\theta)\cos (\phi) \\
	0 & \epsilon_2 & - g_c(\theta)\sin (\phi)\\
	g_c(\theta)\cos (\phi)  & - g_c(\theta)\sin (\phi) & \omega_c
	\end{pmatrix}.
	\end{align}
	
	To obtain the full effective non-Hermitian Hamiltonian of the reduced subspace model, we must similarly reduce the jump operators to their action on this subspace. In the excitonic basis, this is written as,
	\begin{align}\label{eq:H_nh_sub}
	H_i^{(1)}&= - \frac{i}{2} \begin{pmatrix}
	\gamma_{pd}+ 2\Gamma_{th}n(\omega_{vib})  & 0 & 0\\
	0 & \gamma_{pd} + \Gamma_{th}(4n(\omega_{vib}) + 1) & 0\\
	0 & 0&  P_{X_1} + \Gamma_c + 2\Gamma_{th}n(\omega_{vib})
	\end{pmatrix}  - \frac{i}{2} \sum_{\nu , l} \gamma_{\nu, l} \sigma_{\nu, l}^\dagger \sigma_{\nu, l}
	\\ &
	:= 
	- \frac{i}{2} \begin{pmatrix}
	\gamma^{(v)}_{1}  & 0 & 0 \\
	0 & \gamma^{(v)}_{2} & 0 \\
	0 & 0&  \gamma^{(v)}_3
	\end{pmatrix}  - \frac{i}{2} \sum_{\nu , l} \gamma_{\nu, l} \sigma_{\nu, l}^\dagger \sigma_{\nu, l}\nonumber,
	\end{align}
\end{widetext}
	where we have left the polariton decay term out, as it is already in the basis that diagonalises $H$, and thus the vibronic basis on this reduced subspace, making the identification $|F_1\rangle = |v_1\rangle, \, |F_2\rangle = |v_2\rangle, \, |F_3\rangle = |v_c\rangle$, such that $\gamma_{\nu} = \sum_l \gamma_{\nu, l} = \gamma \sum_l F_{\nu, l}$, with $F_{\nu, l} = \langle 1, l| F_\nu\rangle + \langle 2, l| F_\nu\rangle = (\cos (\theta) - \sin(\theta)\langle X_1, l| F_\nu\rangle + (\cos(\theta) + \sin(\theta)) \langle X_2, l | F_\nu \rangle$. We thus have $\gamma_{1/2} = \gamma[(\cos(\theta) \mp \sin(\theta)\cos(\phi) + (\cos(\theta) \pm \sin(\theta))(\pm \sin(\theta) ] $ and similarly $\gamma_2 = \gamma$.
        We can then write this in the vibronic basis, as above,
	\begin{align}\label{eq:Him_sub_vibronic}
	\tilde{H}_i^{(1)}= - \frac{i}{2} \begin{pmatrix}
	\Gamma_1(\phi)  & \Gamma_{12}(\phi) & 0\\
	\Gamma_{21}(\phi) & \Gamma_2(\phi) & 0\\
	0 & 0& \gamma_3
	\end{pmatrix},
	\end{align}

	where $\Gamma_1(\phi) = \gamma^{(v)}_1 \cos^2(\phi) + \gamma^{(v)}_2 \sin^2(\phi) + \gamma_1, \, \Gamma_2 = \gamma^{(v)}_1 \sin^2(\phi) + \gamma^{(v)}_2 \cos^2 (\phi) + \gamma_2, \, \Gamma_{21}(\phi) = \Gamma_{12}(\phi) = (\gamma^{(v)}_1 - \gamma^{(v)}_2)\sin(\phi)\cos(\phi)$. We note that for the parameters considered $\Gamma_{12} \ll \Gamma_1, \Gamma_2, \gamma_3$ and are thus ignored when assigning an effective cooperativity. 

	We thus arrive at a picture that facilitates comparison to the JC Hamiltonian. From the real part, Eq. \eqref{eq:Hr_sub_vibronic}, we can see that there are in this case two vibronic states ($|v_1\rangle$ and $|v_2\rangle$ in Eqs. \eqref{eq:vibronic_states}), which are each coupled to the cavity mode, with coupling strengths that depend on both the excitonic and vibronic mixing angles $\theta, \, \phi$.

	We note that the difference in decay rates between the vibrationally dressed excitonic states is due to the increased thermal dephasing of the state $|X_2, 1, 0, 0\rangle$ relative $|X_1, 0, 0, 0\rangle$, as this rate scales with the vibrational excitation number.
	
	Finally, in analogy to the JC model, we define cavity cooperativities of the vibronic states, as
	\begin{align}\label{eq:DimerCoop_app}
	& C_{1} = \frac{(g_{c}(\theta)\cos(\phi))^2}{\gamma_1^\prime \Gamma_c^\prime }, \nonumber \\ &
	\\&
	C_{2} = \frac{(g_{c}(\theta)\sin(\phi))^2}{\gamma_2^\prime \Gamma_c^\prime }, \nonumber
	\end{align}
	with $\gamma_1^\prime = \gamma_{pd} + \gamma+ 2\Gamma_{th} n(\omega_{vib})$, 
	$\gamma_2^\prime = \gamma_{pd} + \gamma+ \Gamma_{th}(3n(\omega_{vib}) + 1)$, and 
	$\Gamma_c^\prime = P_{X_1} + \Gamma_c + 2\Gamma_{th} n(\omega_{vib})$.
	
\bibliographystyle{apsrev4-1}
\bibliography{Cavity_Paper}

\end{document}